# Properties of FBK UFSDs after neutron and proton irradiation up to $6\cdot 10^{15}$ $n_{eq}/cm^2$

S. M. Mazza[*], E. Estrada, Z. Galloway, C. Gee, A. Goto, Z. Luce, F. McKinney-Martinez, R. Rodriguez, H. F.-W. Sadrozinski, A. Seiden, B. Smithers, Y. Zhao
*SCIPP, Univ. of California Santa Cruz, CA 95064, USA*

V. Cindro, G. Kramberger, I. Mandić, M. Mikuž, M. Zavrtanik
*Jožef Stefan institute and Department of Physics, University of Ljubljana, Ljubljana, Slovenia*

R. Arcidiacono[1,3], N. Cartiglia[1], M. Ferrero[1], M. Mandurrino[1], V. Sola[1,2], A. Staiano[1]
*[1]INFN, [2]Universita' di Torino, [3]Universita' del Piemonte Orientale, Italy*

M. Boscardin[1,2], G.F. Della Betta[2,3], F. Ficorella[1,2], L. Pancheri[2,3], G. Paternoster[1,2]
*[1]Fondazione Bruno Kessler, [2]TIFPA-INFN, [3]Universita' di Trento, Italy*

*Abstract*– The properties of 60-μm thick Ultra-Fast Silicon Detectors (UFSD) detectors manufactured by Fondazione Bruno Kessler (FBK), Trento (Italy) were tested before and after irradiation with minimum ionizing particles (MIPs) from a [90]Sr β-source. This FBK production, called UFSD2, has UFSDs with gain layer made of Boron, Boron low-diffusion, Gallium, carbonated Boron and carbonated Gallium. The irradiation with neutrons took place at the TRIGA reactor in Ljubljana, while the proton irradiation took place at CERN SPS. The sensors were exposed to a neutron fluence of $4\cdot 10^{14}$, $8\cdot 10^{14}$, $1.5\cdot 10^{15}$, $3\cdot 10^{15}$, $6\cdot 10^{15}$ $n_{eq}/cm^2$ and to a proton fluence of $9.6\cdot 10^{14}$ $p/cm^2$, equivalent to a fluence of $6\cdot 10^{14}$ $n_{eq}/cm^2$. The internal gain and the timing resolution were measured as a function of bias voltage at -20°C. The timing resolution was extracted from the time difference with a second calibrated UFSD in coincidence, using the constant fraction method for both.

## 1 - Introduction

The so-called ultra-fast silicon detector (UFSD) will establish a new paradigm for space-time particle tracking [1]. UFSD are thin pixelated n-on-p silicon sensors based on the Low-Gain Avalanche Detector (LGAD) design [2][3][4] developed by the Centro Nacional de Microelectrónica (CNM) Barcelona, in part as a RD50 Common Project [5]. The sensor exhibits moderate internal gain (~5-70) due to a highly doped p+ region just below the n-type implants. First applications of UFSD are envisioned in the upgrades of the ATLAS and CMS experiments at the High-Luminosity Large Hadron Collider (HL-LHC [10]) as reviewed in [11]. In both experiments, the UFSD would be of moderate segmentation (a few mm$^2$) and will face challenging radiation requirements (fluences up to several $10^{15}$ $n_{eq}/cm^2$ and several hundred of MRad).

Previous measurements on different kind of LGAD sensors in beam tests and laboratory before and after irradiation are reported in [7][8][19], these results show that the time resolution of LGADs can be lower than 20 ps. These sets of measurements agree well with the predictions of the simulation program Weightfield2 (WF2) [9]. In all cases the timing resolution has been shown to deteriorate with fluence due

* Corresponding author: simazza@ucsc.edu, telephone (831) 459 1293, FAX (831) 459 5777

to the decreasing value of the gain. This effect is caused by the acceptor removal mechanism [15][20] that decreases the concentration of the active dopant in the gain layer.

In this paper, we report on the performances of 60-µm thick UFSDs produced by Fondazione Bruno Kessler (FBK) before and after a irradiation: the sensors were exposed to a neutron fluence of $4 \cdot 10^{14}$, $8 \cdot 10^{14}$, $1.5 \cdot 10^{15}$, $3 \cdot 10^{15}$, $6 \cdot 10^{15}$ $n_{eq}/cm^2$ and to a proton fluence of $9.6 \cdot 10^{14}$ $p/cm^2$, equivalent to a fluence of $6 \cdot 10^{14}$ $n_{eq}/cm^2$. Devices were irradiated without bias in the JSI research reactor in Ljubljana and at CERN PS IRRAD as explained in Sec. 3.

In Section 2 we will briefly describe the characteristics of the UFSD2 FBK production, already reported in [20]. Section 3 provided the description of the irradiation facilities. In Section 4, a short description of the experimental set-up is presented; details were previously reported in [6], [14] and [19]. In Section 5, we will describe the data analysis including the extraction of the gain and the time resolution, and in Section 6 the results on bias dependence of charge collection and gain, pulse characteristics and timing resolution for a range of fluences will be presented. The performances of the studied UFSDs will be compared.

## 2 – Properties of the FBK UFSDs W1 through W15

As described in more detail in reference [20] 60-µm thick LGAD sensors with different gain layer configurations have been manufactured at FBK. The scope of the production (called UFSD2) was to test the following hypotheses:
1. Gallium might be less prone than Boron to the Watkins mechanism [22, 23].
2. The presence of Carbon atoms might slow down the acceptor removal mechanism, because of the production ion-carbon instead of ion-acceptor complexes.
3. A narrower doping layer with higher initial doping should be less affected by acceptor removal mechanism.

The 5 types of devices are of the following: (i) Boron (B), (ii) Boron low-diffusion (B LD), (iii) Gallium (Ga), (iv) carbonated Boron (B+C), and (v) carbonated Gallium (Ga+C). As seen in Fig. 1 the devices are constituted by an LGAD and PiN (silicon sensor of same thickness with no multiplication layer) in the same dice so to have a correct estimation of the gain by having the same irradiation process for both. It is important to note that Carbon enrichment has been done uniquely in the volume of the gain layer to avoid a sharp increase of the leakage current.

A short summary of the UFSD2 production and irradiation is shown in Table 1: in total 18 wafers were processed, 10 with B-doped and 8 with Ga-doped gain layer. In this paper, we present the response to MIP particles of the most important wafers of the production (wafers 1, 6, 8, 14 and 15). The proton fluence is stated in $n_{eq}/cm^2$ ($6 \cdot 10^{14}$) with the NIEL factor while the real proton fluence is $9.4 \cdot 10^{14} p/cm^2$.

All sensors were studied, however some measurements are missing since a few sensors either broke during handling (eg: W1 sensor for $4 \cdot 10^{14} n_{eq}/cm^2$) or were not tested completely. From the study of the Capacitance over Voltage (CV) curves, the active doping concentration of the gain layer can be calculated from the derivative of $1/C^2$ distribution as explained in [20]: these studies showed that W6 (B+C) retains the highest fraction of active gain layer doping after a given fluence, followed by W15 (Ga+C), then W1 (B LD), W8 (Boron) and lastly W14 (Ga).

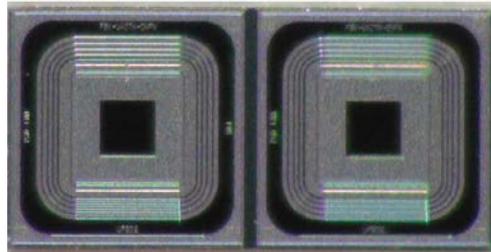

Fig 1: Example of FBK LGAD paired with a PiN with 4 guard-rings.
Each pad is 1x1 mm$^2$ and 60-µm thick.

| Wafer # | Dopant | Gain Dose | N fluence ($n_{eq}/cm^2$) | P fluence ($n_{eq}/cm^2$) |
|---------|--------|-----------|---------------------------|---------------------------|
| 1 | B LD | 0.98 | $8 \cdot 10^{14}$, $1.5 \cdot 10^{15}$, $3 \cdot 10^{15}$ | $6 \cdot 10^{14}$ |
| 6 | B + C | 1.02 | $4 \cdot 10^{14}$, $8 \cdot 10^{14}$, $1.5 \cdot 10^{15}$, $3 \cdot 10^{15}$, $6 \cdot 10^{15}$ | $6 \cdot 10^{14}$ |
| 8 | B | 1.02 | $4 \cdot 10^{14}$, $8 \cdot 10^{14}$, $1.5 \cdot 10^{15}$, $3 \cdot 10^{15}$, $6 \cdot 10^{15}$ | |
| 14 | Ga | 1.04 | $4 \cdot 10^{14}$, $8 \cdot 10^{14}$, $1.5 \cdot 10^{15}$, $3 \cdot 10^{15}$ | $6 \cdot 10^{14}$ |
| 15 | Ga + C | 1.04 | $4 \cdot 10^{14}$, $8 \cdot 10^{14}$, $1.5 \cdot 10^{15}$, $3 \cdot 10^{15}$ | $6 \cdot 10^{14}$ |

Table 1: Wafers tested with charge collection after the irradiation campaign.

## 3 – Irradiations

The UFSD were irradiated without bias in the JSI research reactor of TRIGA type in Ljubljana, which has been used successfully in the past decades to support sensor development [17]. The neutron spectrum and flux are well known and the fluence is quoted in 1 MeV equivalent neutrons per $cm^2$ ($n_{eq}/cm^2$ or shortened $n/cm^2$).

A different set of LGADs was irradiated with protons at the IRRAD CERN irradiation facility [21]. The IRRAD proton facility is located on the T8 beam-line at the CERN PS East Hall where the primary proton beam with a momentum of 24 GeV/c is extracted from the PS ring. In IRRAD, irradiation experiments are performed using the primary protons, prior to reach the beam dump located downstream the T8 beam line. After irradiation, the devices were annealed for 80 min at 60 C. Afterward the devices were kept in cold storage at -20 C.

## 4 – β-telescope setup

The laboratory setup with $^{90}$Sr β-source as well as the readout electronics has been previously described in detail in [6], [14]. Sensors are mounted in single channel readout boards with low noise and high bandwidth trans-impedance amplifiers with a gain of about 10 developed by UCSC. Then the signal is further amplified by a second commercial 2 GHz amplifier of gain 10, which is fast enough to contain most of the LGAD signal that has a frequency spectrum significantly reduced shortly after 1 GHz. The transimpedance of the entire readout is 4700 Ω. The Device Under Test (DUT) is aligned with a trigger sensors that provide also the time reference. The trigger sensor is a HPK UFSD with time resolution of $(17 \pm 1)$ ps, which was measured by pairing two identical UFSDs and evaluating the system time resolution (procedure explained in [6]). The time difference can be found in Fig 2(a), right, the $\sigma_T$ has to be divided by $\sqrt{2}$. Following a trigger pulse, the traces of both trigger and DUT were recorded, with a rate around one Hz with a digital scope with an analogue bandwidth of 2.4 GHz and a digitization step of 50 ps.

The noise was measured as the RMS fluctuation of the base line of the oscilloscope trace. It typically amounts to 1.6mV–2.5mV depending on the type and vertical scale of the oscilloscope. It is important to note that the system is housed in a climate chamber allowing operations of irradiated sensors at lower temperature down to -27°C.

## 5 – Data analysis

The data analysis follows the steps listed in [6]; additional details can be found in [14], [18] and [20]. The digital oscilloscope records the full waveform of both trigger and DUT in each event, so the complete event information is available for offline analysis. The analytic waveforms are then processed offline, the first step is a selection: for a valid trigger pulse, the signal amplitude Pmax of the DUT UFSD should not be saturated by either the scope or the read-out chain. To eliminate the contributions from non-gain events or noise, the time Tmax of the pulse maximum has to fall into a window of 1 ns centered on the trigger pulse. The selected event waveforms are then analyzed to calculate the distribution of the pulse maximum, the collected charge, the gain, the rise and fall time and the time resolution.

The collected charge is measured by integrating the voltage pulse and dividing it by the trans-impedance of the amplifier. The distribution is then fitted to a Landau curve and the MPV is divided by the MPV of the collected charge of a PiN with the same thickness and neutron fluence to calculate the gain.

The time of arrival of a particle is defined with the constant fraction discriminator (CFD) method [14] [18], where the time of arrival of the pulse is evaluated at a % level of the pulse maximum. Due to the oscilloscope digitization steps, the time of arrival at a specific CFD fraction is evaluated with a linear interpolation. This method offers a very efficient correction to the time walk effect. The CFD value can be optimized for every bias voltage and fluence to minimize the time resolution, a procedure that is necessary since both the pulse shape and the noise contributions change with fluence. The time resolution of the system is derived from the sigma of a Gaussian fit to the time difference Δt between the DUT and the trigger (Fig 2(a), left), both corrected for time walk with its proper CFD level. The resolution σt of the DUT is then calculated by removing in quadrature the trigger contribution.

The normalized average pulse shape for both sensors before and after irradiation for W6 (carbonated Boron) and W8 (Boron only) sensors can be seen in Fig. 2(b), indicating an increase of the slope of the rising edge for both. The faster rise time of the irradiated sensors is caused by the increased bias voltage which is 280V before irradiation and 600V after irradiation. The effect is more prominent in W8 since the doping concentration of the gain layer is fairly low after irradiation and the pulse mostly comes from the bulk with no multiplication.

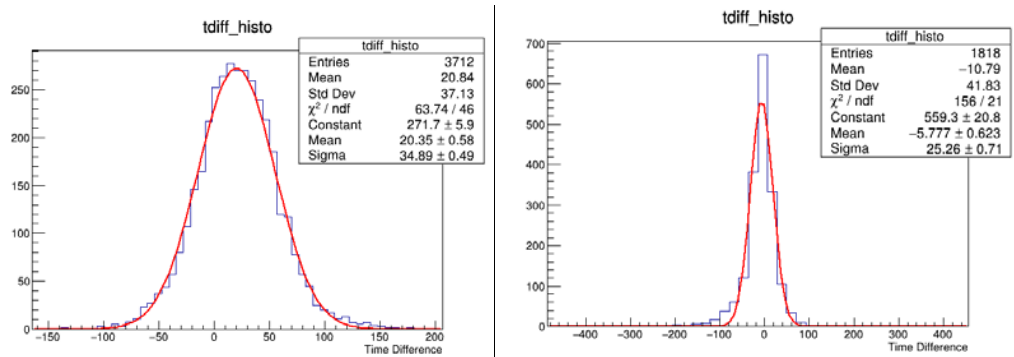

Fig 2(a): Left, Gaussian fit of the time difference between DUT (FBK UFSD2 W6 pre-rad) and trigger
Right, Gaussian of the time difference between two identical trigger LGADs

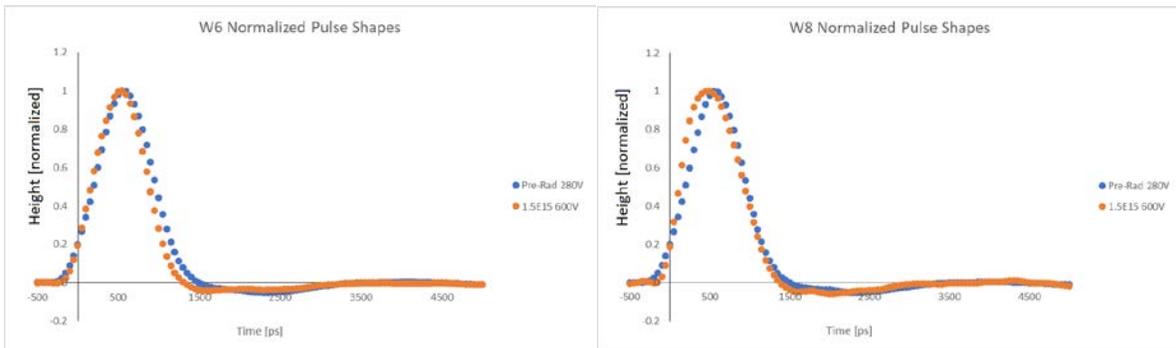

Fig 2(b): Average pulse shape normalized to pulse maximum
for W6 (carbonated Boron) (left) and W8 (Boron only) (right)
before and after 1.5E15 $n_{eq}/cm^2$ irradiation.

# 6 - Results

In this section, we present a brief description of the most important results obtained so far

## 6.1 Gain vs bias for different fluences

As observed in previous measurements [14], the value of the gain of UFSD decreases with fluence and can in part be recovered by increasing the bias voltage applied to the sensor, as shown in Fig. 3. For UFSD2 sensors, this decrease is less or more pronounced for the different types of gain layer doping. Fig. 4 shows the maximum operating voltage for sensor after irradiation, defined as the maximum bias voltage that can be applied to the sensor before breakdown. Some sensors showed premature noisy behavior and high current (for example W15 at $1.5 \cdot 10^{15}$ $n_{eq}/cm^2$) so had to be operated at lower voltages. All measurements for irradiated sensors were taken at -20 C, while for not irradiated sensors at 20C.

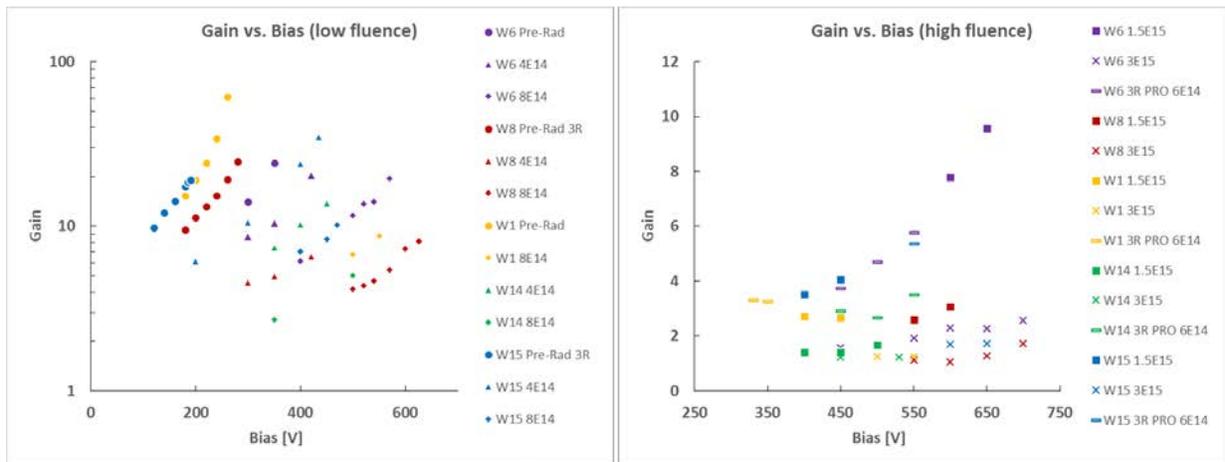

Fig 3: Gain as a function of bias voltage applied for low (left) and high (right) irradiation fluence.
W1 (Boron LD) W6 (Boron Carbon) W8 (Boron) W14 (Gallium) W15 (Gallium Carbon)

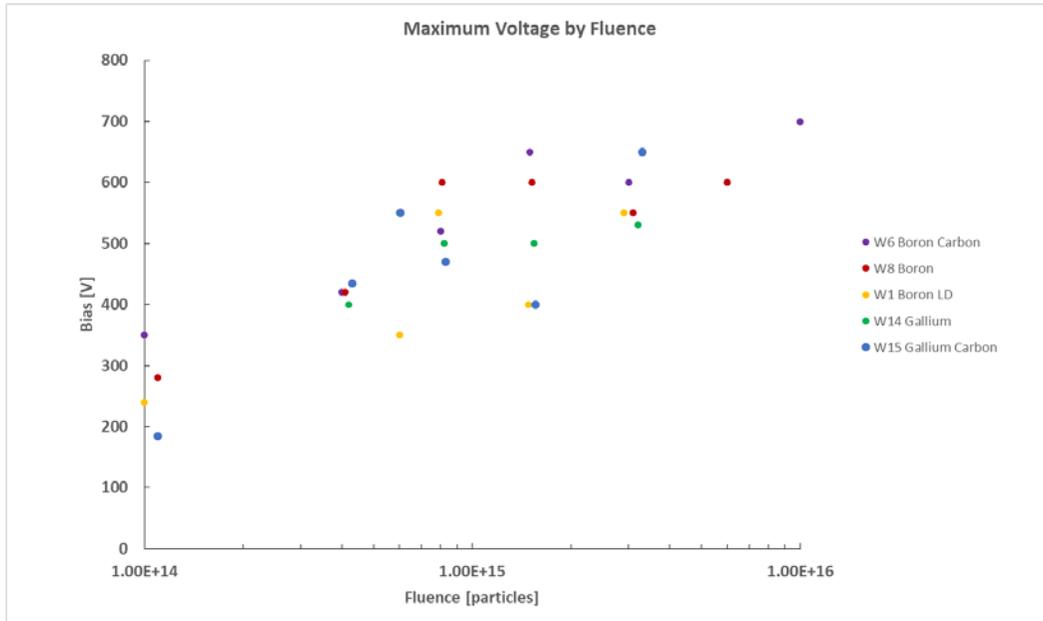

Fig 4: Maximum operating voltage for different fluences.
W1 (Boron LD) W6 (Boron Carbon) W8 (Boron) W14 (Gallium) W15 (Gallium Carbon)
(The values before irradiation are shown at a fluence of 1E14 n/cm$^2$)

6.2 Rise time

An important parameter determining the time resolution is the rise time: for un-irradiated sensors, the rise time is determined by the electron drift time, while for irradiated sensors it becomes shorter as the multiplication mechanism moves from the gain layer to the bulk and the bias voltage increases [14]. The dependence of the rise time (10 – 90%) upon the bias voltage is shown in Fig 5. For all sensors, the rise time decreases with progressive irradiation.

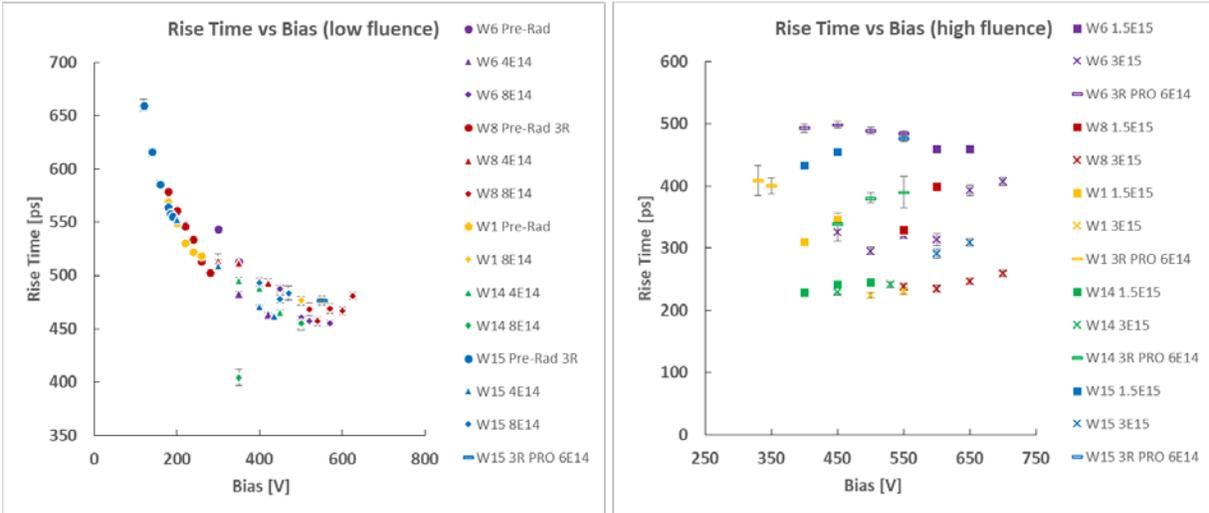

Fig 5: Rise time (10 - 90%) for low (left) and high (right) irradiation fluence as a function of bias voltage.
W1 (Boron LD) W6 (Boron Carbon) W8 (Boron) W14 (Gallium) W15 (Gallium Carbon)

In Fig. 6, the rise time as a function of gain is shown. For very highly irradiated sensors which exhibit low gain, the rise time increase with increasing gain, before decreasing slightly. This can be explained by the fact that the contribution of the original drifting charge has a faster rise time than the drifting holes from the multiplication process, and thus have a larger influence on the pulse shape at low gain.

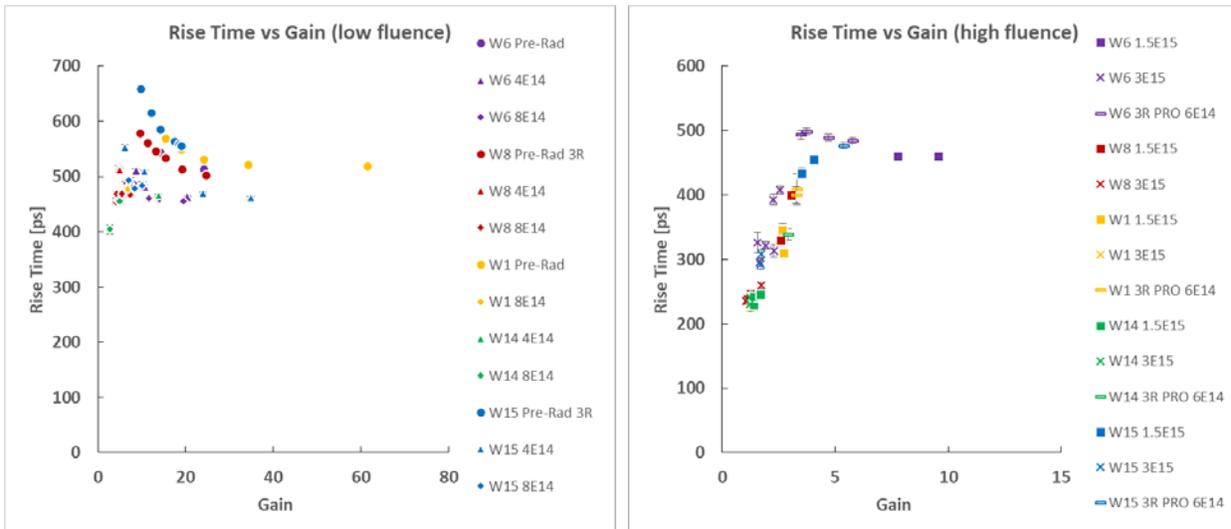

Fig 6: Rise time (10 - 90%) for low (left) and high (right) irradiation fluence as a function of gain.
W1 (Boron LD) W6 (Boron Carbon) W8 (Boron) W14 (Gallium) W15 (Gallium Carbon)

6.3 Time resolution

The time resolution of UFSD of all wafers is roughly the same before irradiation (Fig. 7). This information is very important as it indicates that neither the addition of Carbon nor the presence of Gallium in the gain layer can cause a degradation of performances. Comparing the performances of the different wafers as a function of fluence, evaluated at 20 % CFD, we see that the time resolution of W6 (B+C) is better to those of all other wafers starting from a fluence of $8 \cdot 10^{14}$ $n_{eq}/cm^2$. W6 is the only wafer able to reach a time resolution of sub 40 ps after $1.5 \cdot 10^{15}$ $n_{eq}/cm^2$.

Fig. 8 shows the time resolution as a function of a CFD scan from 10 % to 99 % for W6 and W8 sensors. The difference at a fluence of 1.5E15 is notable, for W6 the optimal CFD is still around 20% while for W8 it increases to 44%. This shows that for W6 a large fraction of the original gain layer is still active, as already seen in Fig 3 and in the results from the C-V measurements [20]. In Table 2 the best ("optimized") CFD and the range (for 10% of time resolution variation from the minimum) for each sensor and fluence is shown. As a function of fluence, the value of CFD needs to be moved to higher values since the Jitter contribution is smaller at higher CFD settings.
If Fig. 9 is the time resolution for the optimized CFD threshold, the time resolution performance of the all wafers improves, especially at higher fluences. Comparing results in Fig. 9 with results in Fig. 7 it can be seen that, as example, the time resolution for W8 at $1.5 \cdot 10^{15}$ $n_{eq}/cm^2$ goes from 48 ps (for 20% CFD) to 38 ps (for 44% CFD), at $3 \cdot 10^{15}$ $n_{eq}/cm^2$ it goes from 66.5 ps (for 20% CFD) to 48.9 ps (for 68% CFD).
Also UFSD from wafers W14 (Ga), W15 (Ga + C) and W1 (B LD) have a time resolution around 50 ps using a value of CFD around 50%-70% for the higher fluences while W6 (B+C) has comparable or better performance with a lower CFD value.

At very high fluence ($6 \cdot 10^{15} n_{eq}/cm^2$), the multiplication layer is completely de-activated for every gain layer type for both W6 and W8 requiring a very high CFD value (70-75%) to get a time resolution around 50 ps (not shown in the plots). All the values showed in the plots are measured at -20 C for irradiated sensors and at room temperature for not irradiated sensors. Some of the sensors were also tested at -27 C but no performance improvements were observed.

| W# | Dopant | Dose | Neutron | | | | | | Proton | |
|---|---|---|---|---|---|---|---|---|---|---|
| 1 | B LD | 0.98 | 0 | $4 \cdot 10^{14}$ | $8 \cdot 10^{14}$ | $1.5 \cdot 10^{15}$ | $3 \cdot 10^{15}$ | $6 \cdot 10^{15}$ | $6 \cdot 10^{14}$ | fluence ($n_{eq}/cm^2$) |
| | | | 5 | | 14 | 52 | 70 | | 38 | Optimal CFD % |
| | | | 2-10 | | 8-25 | 31-68 | 38-96 | | 18-48 | CFD range |
| 6 | B + C | 1.02 | 0 | $4 \cdot 10^{14}$ | $8 \cdot 10^{14}$ | $1.5 \cdot 10^{15}$ | $3 \cdot 10^{15}$ | $6 \cdot 10^{15}$ | $6 \cdot 10^{14}$ | fluence ($n_{eq}/cm^2$) |
| | | | 11 | 9 | 16 | 13 | 28 | 70 | 16 | Optimal CFD % |
| | | | 6-18 | 5-16 | 9-32 | 8-20 | 20-45 | 48-89 | 9-25 | CFD range |
| 8 | B | 1.02 | 0 | $4 \cdot 10^{14}$ | $8 \cdot 10^{14}$ | $1.5 \cdot 10^{15}$ | $3 \cdot 10^{15}$ | $6 \cdot 10^{15}$ | $6 \cdot 10^{14}$ | fluence ($n_{eq}/cm^2$) |
| | | | 8 | 23 | 15 | 44 | 68 | 75 | | Optimal CFD % |
| | | | 3-12 | 9-37 | 9-26 | 23-60 | 28-83 | 48-96 | | CFD range |
| 14 | Ga | 1.04 | 0 | $4 \cdot 10^{14}$ | $8 \cdot 10^{14}$ | $1.5 \cdot 10^{15}$ | $3 \cdot 10^{15}$ | $6 \cdot 10^{15}$ | $6 \cdot 10^{14}$ | fluence ($n_{eq}/cm^2$) |
| | | | 11 | 11 | 26 | 77 | 58 | | 32 | Optimal CFD % |
| | | | 5-20 | 5-20 | 18-31 | 48-96 | 45-73 | | 26-34 | CFD range |
| 15 | Ga + C | 1.04 | 0 | $4 \cdot 10^{14}$ | $8 \cdot 10^{14}$ | $1.5 \cdot 10^{15}$ | $3 \cdot 10^{15}$ | $6 \cdot 10^{15}$ | $6 \cdot 10^{14}$ | fluence ($n_{eq}/cm^2$) |
| | | | 12 | 10 | 20 | 32 | 54 | | 23 | Optimal CFD % |
| | | | 6-19 | 6-28 | 14-30 | 23-42 | 48-70 | | 12-34 | CFD range |

Table 2: Optimal CFD for several sensors and fluence. The CFD range is written as min-max CFD to have a 10% variation on the time resolution at the minimum.

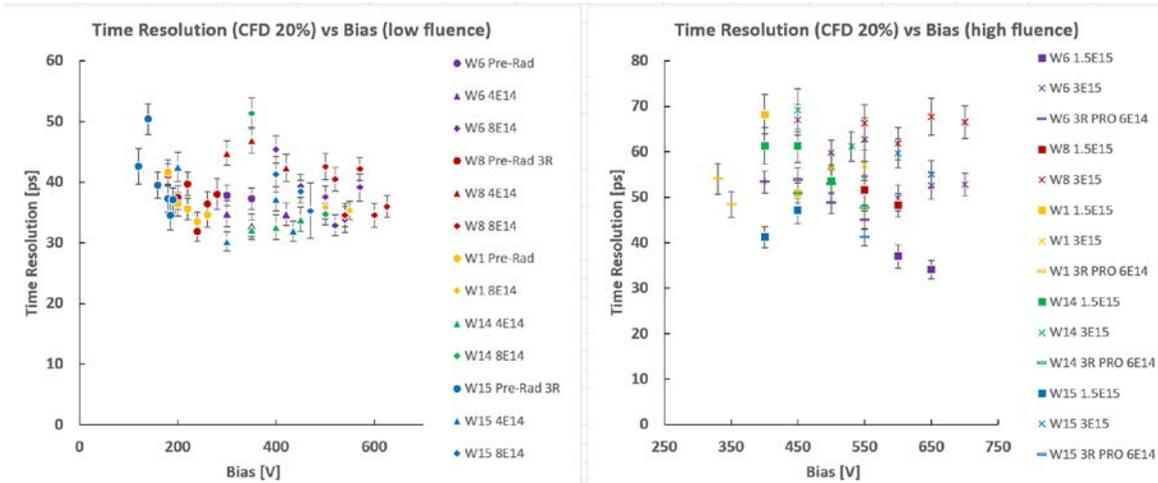

Fig 7: Time resolution for low (left) and high (right) irradiation fluence as a function of bias voltage for a CFD threshold of 20 %. W1 (Boron LD) W6 (Boron Carbon) W8 (Boron) W14 (Gallium) W15 (Gallium Carbon)

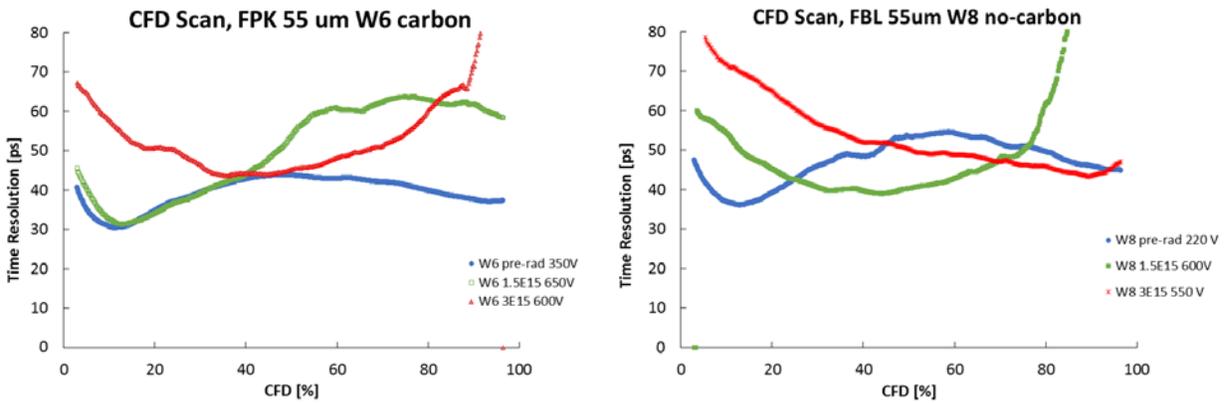

Fig 8: CFD scan at optimum operating voltage for FBK sensors W6 (Boron Carbon) (left) and W8 (Boron) (right).

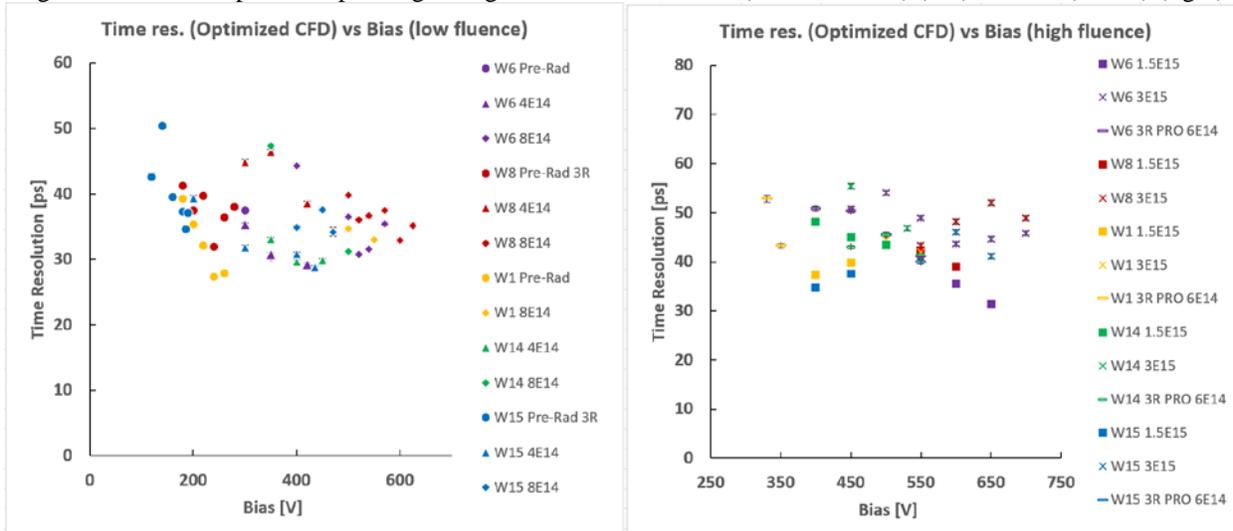

Fig 9: Time resolution for low (left) and high (right) irradiation fluence as a function of bias voltage for a CFD threshold optimized for the optimal time resolution.
W1 (Boron LD) W6 (Boron Carbon) W8 (Boron) W14 (Gallium) W15 (Gallium Carbon)

The CFD-optimized time resolutions as a function of fluence are presented in Fig. 10.

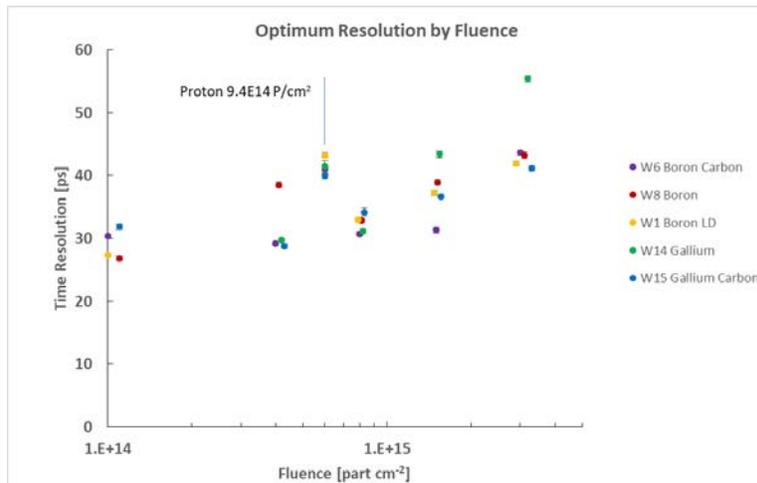

Fig 10: Time resolution as a function of fluence for FBK sensors. The wafer, although all irradiated at the same fluence, are represented at slightly different fluences for displaying purposes.
The proton data is showed at $6 \cdot 10^{14}$ $n_{eq}$/cm$^2$ indicated by a line in the plot, the proton fluence is $9.4 \cdot 10^{14}$ P/cm$^2$.
W1 (Boron LD) W6 (Boron Carbon) W8 (Boron) W14 (Gallium) W15 (Gallium Carbon)
(The values before irradiation are shown at a fluence of 1E14 n/cm$^2$)

6.4 Leakage current and noise

In Fig. 11 the leakage current of the sum of the sum of pad (size of 1x1 mm$^2$) and guard ring is shown, the current was estimated during the charge collection measurement. Before irradiation carbonated wafers (in particular W6) show higher current than non-carbonated wafers most likely due to the defect caused by carbon implantation. However, after irradiation the leakage current is similar for all detectors. The current after irradiation changes linearly with temperature, roughly a factor 2 every 7 C, as expected for irradiated Silicon. The contribution of the guard ring to the total current is 10-20 %, since the current produced by the irradiated bulk is multiplied by the gain layer. The noise is shown on Fig. 12, it is roughly constant around 1.5 mV with spikes from some of the detectors.

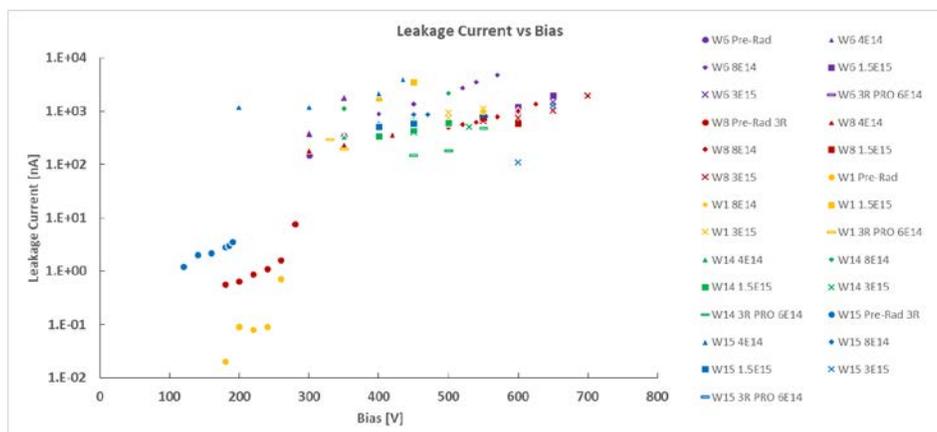

Fig 11: Leakage current for all FBK wafer before and after irradiation. Although W6 (Carbonated Boron) has a higher leakage current before irradiation it becomes similar to the other wafer after radiation damage.
W1 (Boron LD) W6 (Boron Carbon) W8 (Boron) W14 (Gallium) W15 (Gallium Carbon)

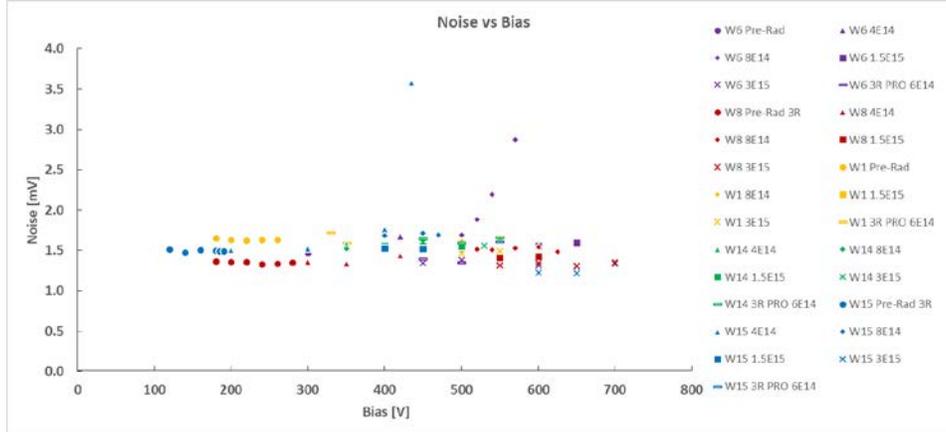
Fig 12: Noise on the oscilloscope as a function of Bias.
W1 (Boron LD) W6 (Boron Carbon) W8 (Boron) W14 (Gallium) W15 (Gallium Carbon)

## 7 - Conclusions

Several UFSD from Fondazione Bruno Kessler (FBK) of 60-µm thickness were tested using a $^{90}$Sr β-telescope. The sensors were evaluated before and after neutron irradiation up to $3 \cdot 10^{15}$ n/cm$^2$ and for proton irradiation of $6 \cdot 10^{14}$ n$_{eq}$/cm$^2$. The operating voltage of the sensors ranges from 350 V (new) to 700 V ($3 \cdot 10^{15}$ n/cm$^2$ neutron irradiation).

All wafers have similar performance in terms of time resolution before irradiation. Post irradiation, carbonated boron sensors from wafer W6 (Boron Carbon) have equal or better time resolution and gain than the others using a fixed CFD of 20%. These results are in line with the preliminary studies done on the CV of the sensors in [20]. Some of the performance of the other type of sensors can be regained after irradiation by increasing the CFD threshold. Even with CFD optimization W6 maintains a CFD between 20 % and 35 % until $3 \cdot 10^{15}$ n/cm$^2$. Sensor with not carbonated boron (W1, W8) have similar performance to W6 sensor if the CFD threshold is increased up to 60% for $3 \cdot 10^{15}$ n/cm$^2$ of neutron fluence. Sensor with Gallium (W14, W15) doping instead show equal or worse performance than Boron sensors after irradiation, carbonated Gallium sensors (W15) are better than Gallium only sensors (W14).

The optimized time resolution of carbonated sensor is ~30 ps when new, this value is roughly maintained until $1.5 \cdot 10^{15}$ n/cm$^2$ of neutron radiation damage. For the higher fluence of $3 \cdot 10^{15}$ n/cm$^2$ ($1 \cdot 10^{16}$ n/cm$^2$) the time resolution increases to ~45 (~50). The results from the proton irradiation will be reported in the future.

## 8 - Acknowledgements

We acknowledge the contribution to this paper by the FBK team and the expert technical help by the SCIPP technical staff.
Part of this work has been performed within the framework of the CERN RD50 Collaboration.
The work was supported by the United States Department of Energy, grant DE-FG02-04ER41286. Part of this work has been financed by the European Union's Horizon 2020 Research and Innovation funding program, under Grant Agreement no. 654168 (AIDA-2020) and Grant Agreement no. 669529 (ERC UFSD669529), and by the Italian Ministero degli Affari Esteri and INFN Gruppo V.

# 9 - References

[1] H.F.-W. Sadrozinski, A. Seiden and N. Cartiglia, "4D tracking with ultra-fast silicon detectors", 2018 Rep. Prog. Phys. 81 026101.

[2] G. Pellegrini, et al., "Technology developments and first measurements of Low Gain Avalanche Detectors (LGAD) for high energy physics applications", Nucl. Instrum. Meth. A765 (2014) 24.

[3] H.F.-W. Sadrozinski, et al., "Ultra-fast silicon detectors", Nucl. Instrum. Meth. A831 (2016) 18.

[4] M. Carulla et al., "First 50 µm thick LGAD fabrication at CNM", 28th RD50 Workshop, Torino, June 7th 2016, https://agenda.infn.it/getFile.py/access?contribId=20&sessionId=8&resId=0&materialId=slides&confId=11109.

[5] RD50 collaboration, http://rd50.web.cern.ch/rd50/.

[6] N. Cartiglia et al., "Beam test results of a 16 ps timing system based on ultra-fast silicon detectors", Nucl. Instrum. Meth. A850, (2017), 83–88.

[7] N. Cartiglia et al, "Performance of Ultra-Fast Silicon Detectors", JINST 9 (2014) C02001.

[8] N. Cartiglia, "Design optimization of ultra-fast silicon detectors", Nucl. Instrum. Meth. A796 (2015) 141-148.

[9] F. Cenna, et al., "Weightfield2: A fast simulator for silicon and diamond solid state detector", Nucl. Instrum. Meth. A796 (2015) 149; http://personalpages.to.infn.it/~cartigli/Weightfield2/Main.html.

[10] HL-LHC, http://dx.doi.org/10.5170/CERN-2015-005.

[11] L. Gray, "4D Trackers", at "Connecting the dots", Paris 2017, https://indico.cern.ch/event/577003/contributions/2476434/attachments/1422143/2180715/20170306_LindseyGray_CDTWIT.pdf.

[12] G. Kramberger et al.: "Radiation hardness of thin LGAD detectors", TREDI 2017, https://indico.cern.ch/event/587631/contributions/2471705/attachments/1414923/2165831/RadiationHardnessOfThinLGAD.pdf.

[13] J. Lange et al "Gain and time resolution of 45 µm thin LGAD before and after irradiation up to a fluence of $10^{15}$ neq/cm$^2$", JINST **12** P05003

[14] Z. Galloway et al, "Properties of HPK UFSD after neutron irradiation up to 6e15 n/cm$^2$" arXiv:1707.04961.

[15] G. Kramberger et al., "Radiation effects in Low Gain Avalanche Detectors after hadron irradiations", JINST 10 P07006, 2015.

[16] G. Kramberger, et al., "Effective trapping time of electrons and holes in different silicon materials irradiated with neutrons, protons and pions", Nucl. Instrum. Methods Phys. Res. A 481 (2002) 297–305

[17] Snoj, G. ˇZerovnik and A. Trkov, "Computational analysis of irradiation facilities at the JSI TRIGA reactor", Appl. Radiat. Isot. 70 (2012) 483.

[18] Y Zhao, UCSC Senior Thesis 2017, https://drive.google.com/drive/folders/0ByskYealR9x7bFY1ZS1pZW9SRWs.

[19] Y. Zhao et al, "Comparison of 35 and 50 μm thin HPK UFSD after neutron irradiation up to 6*10^15 neq/cm2" arXiv:1803.02690.

[20] M.Ferrero et al, "Radiation resistant LGAD design", arXiv:1802.01745, 10.1016/j.nima.2018.11.121

[21] B. Gkotse, et al. Irradiation Facilities at CERN [http://cds.cern.ch/record/2288578/files/AIDA-2020-POSTER-2017-004.pdf].

[22] G. Kramberger, et al., Radiation hardness of gallium doped low gain avalanche detectors, Nuclear Instruments and Methods in Physics Research Section A: Accelerators, Spectrometers, Detectors and Associated Equipment 898 (2018) 53 { 59. doi:https://doi.org/10.1016/j.nima.2018.04.060. URL http://www.sciencedirect.com/science/article/pii/S0168900218305771

[23] A. Khan, et al., Strategies for improving radiation tolerance of Si space solar cells, Solar Energy Materials and Solar Cells 75 (1) (2003) 271 { 276, pVSEC 12 Part II. doi:10.1016/S0927-0248(02) 00169-1. URL http://www.sciencedirect.com/science/article/pii/S0927024802001691

# Appendix A. – Time resolution vs bias voltage per wafer

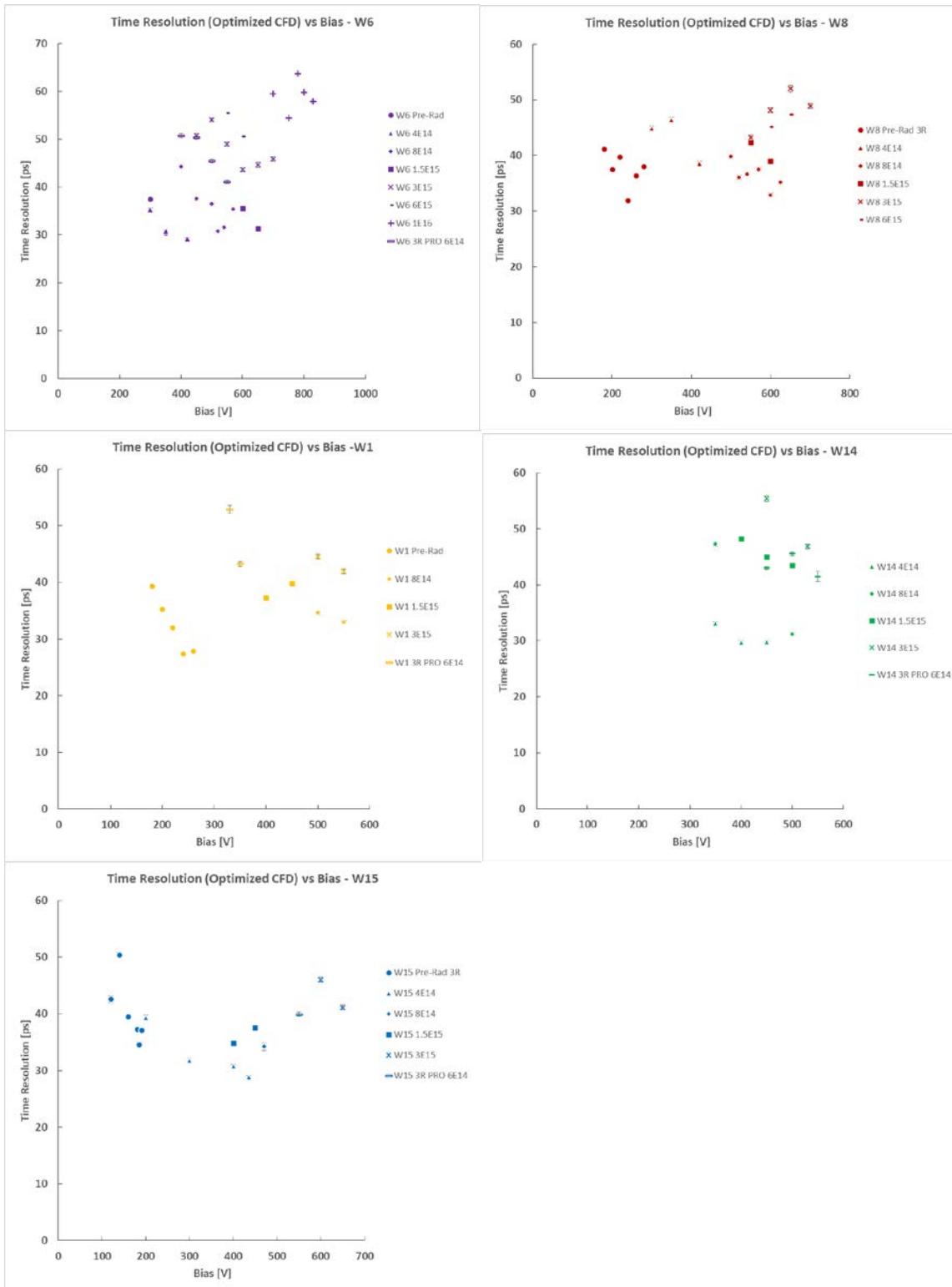

# Appendix B. – Time resolution vs bias voltage per fluence

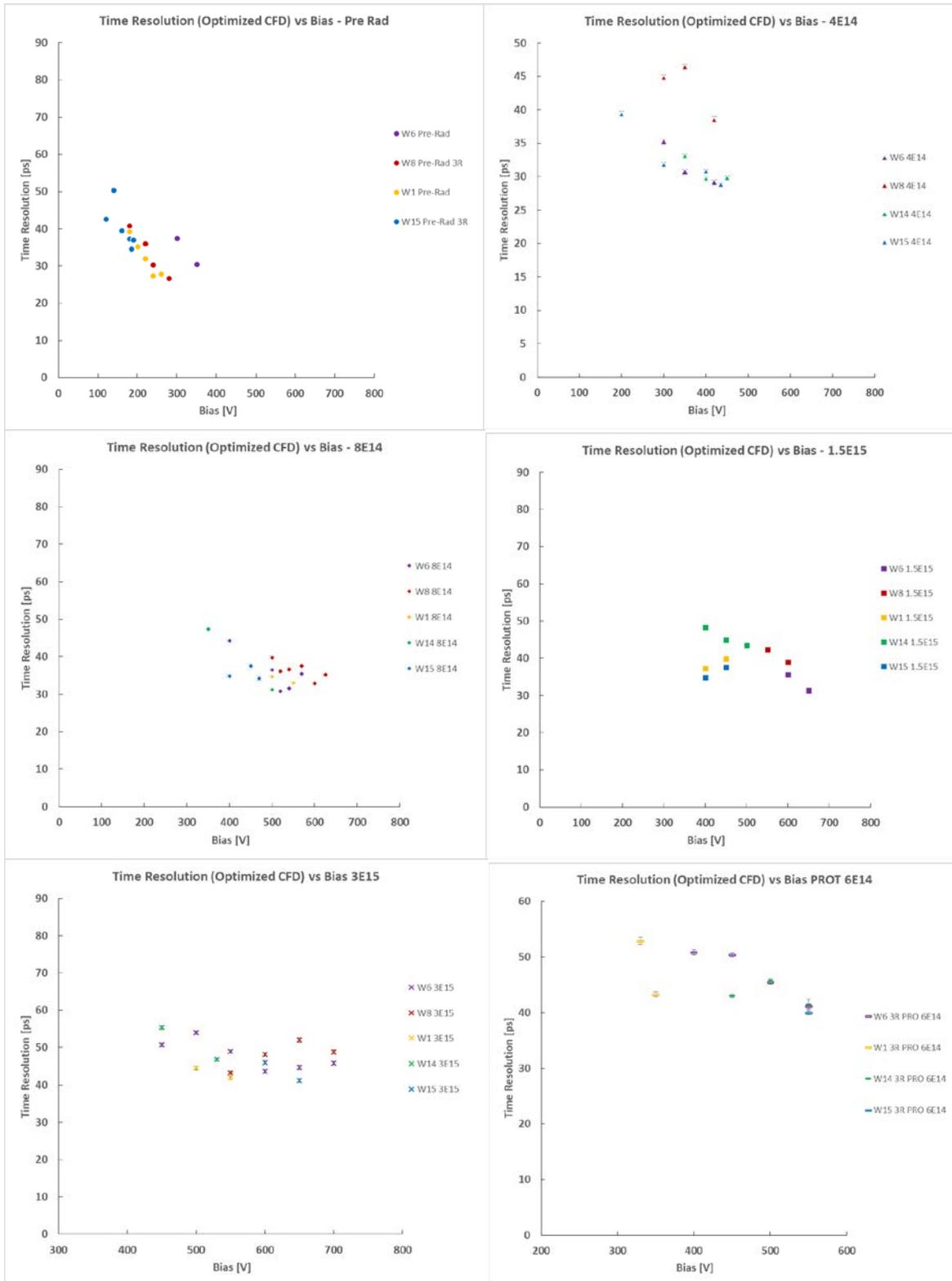

## Appendix C – Gain calculation

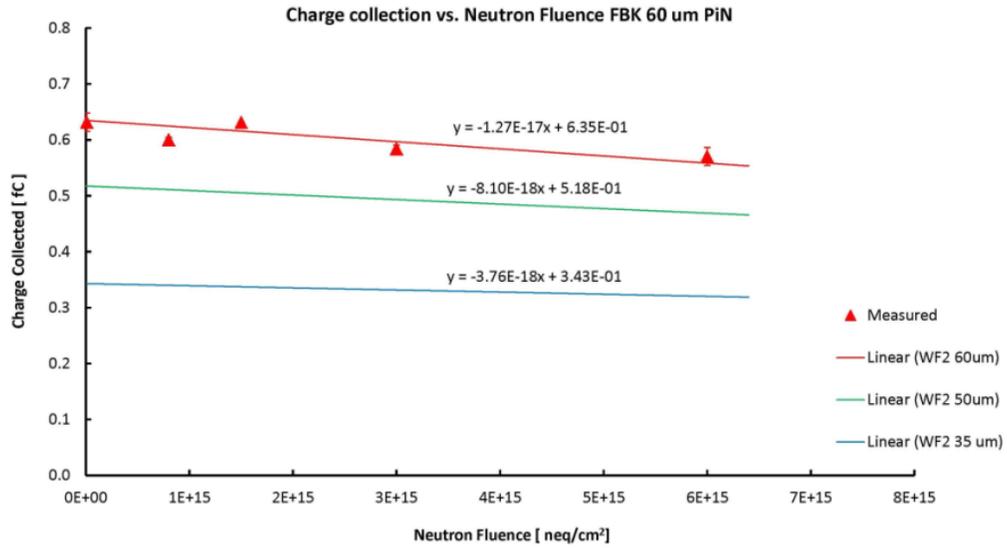

Fig A.1: Collected charge for 60um FBK PiN diodes (red triangles) for different irradiation fluences. The colored lines shows the WF2 simulation with parameters tuned to reproduce the 60um PiN data. The simulation shown is for 35um and 50um PiN diodes.